\documentclass[namedreferences]{SolarPhysics}
\usepackage[optionalrh]{spr-sola-addons} % For Solar Physics 
\usepackage{graphicx}        % For eps figures, newer & more powerful
\usepackage{color}           % For color text: \color command
\usepackage{url}             % For breaking URLs easily trough lines
\begin{document}

\begin{opening}

\title{Correlation functions of small-scale fluctuations of the interplanetary magnetic field}

\author{Z.~\surname{N{\'e}meth}$^{1}$\thanks{These authors contributed equally to this work.}\sep
        G.~\surname{Facsk{\'o}}$^{2}$$^{3}$\footnotemark[1]\sep
        E.~A.~\surname{Lucek}$^{4}$      
       }
\runningauthor{N{\'e}meth, Facsk{\'o}, and Lucek}
\runningtitle{Correlation functions of small-scale IMF fluctuations}

   \institute{$^{1}$ KFKI Research Institute for Particle and Nuclear Physics, H-1525 Budapest, P.O.Box 49, Hungary
                     email: \url{nemeth@rmki.kfki.hu} \\ 
              $^{2}$ HAS Geodetic and Geophysical Research Institute, H-9401 Sopron, P.O.Box 5, Hungary                
                     email: \url{gfacsko@ggki.hu} \\
              $^{3}$ Laboratoire de Physique et Chimie de l'Environnement et de l'Espace/CNRS, 3A, Avenue de la Recherche Scientifique, 45071 Orl\'eans cedex 2, France \\
              $^{4}$ The Blackett Laboratory, Imperial College London, Prince Consort Road, London SW7 2BW, UK
                     email: \url{e.lucek@imperial.ac.uk} \\
              }

\begin{abstract}
The Interplanetary Magnetic Field shows complex spatial and temporal variations. Single spacecraft measurements reveal only a one dimensional section of this rich four dimensional phenomenon. Multi-point measurements of the four Cluster spacecraft provide a unique tool to study the spatiotemporal structure of the field. Using Cluster data we determined three dimensional correlation functions of the fluctuations. By means of the correlation function one can describe and measure field variations. Our results can be used to verify theoretical predictions, to understand the formation and nature of solar wind turbulence. We found that the correlation length varies over almost six orders of magnitude. The IMF turbulence shows significant anisotropy with two distinct populations. In certain time intervals the ratio of the three axes of the correlation ellipse is 1/2.2/6 while in the remaining time we found extremely high correlation along one axis. We found favoured directions in the orientation of the correlation ellipsoids.
\end{abstract}


\end{opening}

%-------------------------------------------------

\begin{article}

\section{Introduction and motivation}
\label{sec:intro} 

The Interplanetary Magnetic Field (IMF) shows complex spatial and temporal variations which have been of much interest for a long time now. The phenomenon is often referred to as the turbulence of the solar wind. It is an important factor in the energy and momentum transport of the plasma, influences the formation and dynamics of shock waves and other discontinuities. The turbulence of the solar wind has its influence even on the propagation of cosmic ray particles \cite{bieber96:_domin_two_dimen_solar_wind} or on the distribution and mean free path of pickup ions \cite{nemeth00:_fluct_of_helios_magnet_field}. Every detailed model of the solar wind must account for these strong fluctuations. 

The magnetic variations of the solar wind  are supposed to be the superposition of two types of fluctuations: the so called ``2D-turbulence'' and slab variations. In the model of slab disturbances the fluctuation wave vector is aligned with the mean magnetic field. This model has a physical motivation in terms of Alfv\'en waves propagating along the mean
field. In the 2D model however the turbulence wave vectors are predominantly in the plane perpendicular to the mean magnetic field \cite{matthaeus90:_eviden_for_presen_of_quasi,bieber96:_domin_two_dimen_solar_wind}. An interpretation of the 2D model using resonant three wave interactions was provided by \inlinecite{shebalin83:_anisot_in_mhd_turbul_due}. More recently an alternative explanation was provided based on plasma mixing \cite{nemeth06:_plasm_mixin_as_cause_of}.

The solar wind turbulence is known to be anisotropic and the contour plot of the two-dimensional correlation of the fluctuations as a function of distance parallel and perpendicular to the mean magnetic field shows a ``Maltese cross'' shape \cite{matthaeus90:_eviden_for_presen_of_quasi,matthaeus05:_spatial_correl_of_solar_wind}. The Ulysses mission made it possible to study the turbulence in the high latitude solar wind as well \cite{bigazzi06:_small_scale_anisot_and_inter}. The anisotropy is present in the fast and slow solar wind alike \cite{dasso05:_anisot_in_fast_and_slow}. 

Single spacecraft measurements \cite{matthaeus90:_eviden_for_presen_of_quasi,bieber96:_domin_two_dimen_solar_wind,dasso05:_anisot_in_fast_and_slow,bigazzi06:_small_scale_anisot_and_inter} however only reveal a one dimensional section of this rich 3+1 dimensional phenomenon, which means that the spatial structure can only be estimated using strong assumptions. To avoid this limitation  spacecraft couples \cite{richardson01:_plasm_and_magnet_field_correl} were used at first. The multi-point measurements of the four Cluster spacecraft provide an even better possibility to study the spatial-temporal structure of the field \cite{horbury00:_clust_ii_analy_of_turbul,osman07:_multis_measur_of_anisot_correl,osman09:_multi_spacec_measur_of_anisot,osman09:_quant_estim_of_slab_and}.

The principal tool generally used to describe and measure field fluctuations is the two point correlation function. Our goal is to determine the complete three dimensional (3D) correlation functions of small scale IMF fluctuations. Previous studies only determined sections of the complete 3D function, used sometimes questionable symmetry assumptions and/or used relatively small data samples. In this paper we analyze more than 200 hours of data, deal with three axis anisotropy and reconstruct the complete 3D correlation function from its measured sections. Using the correlation function we are able to calculate for example three axes correlation ellipsoids, anisotropy and correlation length ratios. Our results can be used to verify theoretical predictions, to investigate turbulence and intermittency as well as the transport properties of the solar wind. 

In Section~\ref{sec:obs} we present the observations, in Section~\ref{sec:tech} the methods are described and finally we summarize our results in Section~\ref{sec:disc}. 

\section{Observations}
\label{sec:obs}

\begin{figure}
\centerline{\includegraphics[width=0.45\textwidth,clip=]{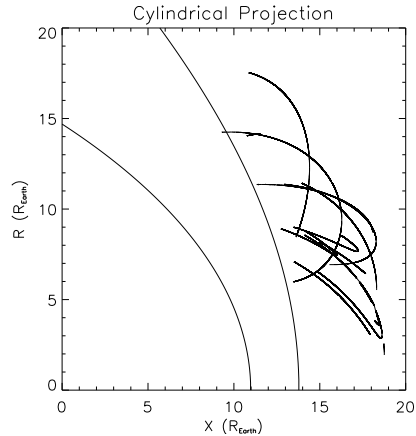}
\caption{The cylindrical projection of Cluster spacecraft orbits during our observations in  GSE  system, where X and $R=\sqrt{Y_{GSE}^2+Z_{GSE}^2}$ are measured in $R_{Earth}$ units. A model magnetopause (left curve) and bow shock (right curve) are also shown for reference.}\label{fig:fig01}}
\end{figure}

\begin{table}
\begin{tabular}{ccccc}
\multicolumn{2}{c}{start date and time} & \multicolumn{2}{c}{end date and time} & duration \\
(yyyymmdd) & (hh:mm:ss) & (yyyymmdd) & (hh:mm:ss) & (hh:mm) \\
\hline
20030206 & 00:00:00 & 20030206 & 18:00:00 & 18:00 \\
20030211 & 02:00:00 & 20030211 & 10:00:00 & 08:00 \\
20030212 & 19:00:00 & 20030213 & 00:00:00 & 05:00 \\
20030217 & 03:00:00 & 20030218 & 00:00:00 & 21:00 \\
20030219 & 23:10:00 & 20030220 & 00:00:00 & 00:50 \\
20030220 & 08:00:00 & 20030221 & 00:00:00 & 16:00 \\
20030222 & 09:00:00 & 20030223 & 00:00:00 & 15:00 \\
20030223 & 01:00:00 & 20030223 & 03:15:00 & 02:15 \\
20030224 & 16:00:00 & 20030225 & 00:00:00 & 08:00 \\
20030227 & 14:00:00 & 20030227 & 21:00:00 & 07:00 \\
20030228 & 00:00:00 & 20030228 & 03:00:00 & 03:00 \\
20040316 & 13:00:00 & 20040317 & 02:30:00 & 13:30 \\
20050402 & 09:00:00 & 20050403 & 13:00:00 & 28:00 \\
20060118 & 17:30:00 & 20060118 & 19:30:00 & 02:00 \\
20070218 & 00:00:00 & 20070218 & 17:00:00 & 17:00 \\
20070219 & 22:00:00 & 20070221 & 00:00:00 & 26:00 \\
20080420 & 16:00:00 & 20080421 & 12:00:00 & 20:00 \\ 
\end{tabular}
\caption{The intervals studied. During these intervals the Cluster spacecraft dwelt inside the solar wind.}
\label{tab:intervalls}
\end{table}

For this study we have used 1s averaged time resolution Cluster II Flux Gate Magnetometer (FGM) data \cite{balogh01:_clust_magnet_field_inves}. The spacecraft separation was in the range of several hundreds to several thousands of kms during the observations. The elongated ellipse orbit of Cluster spacecraft samples the solar wind in February, March and April and allows interplanetary magnetic field (IMF) measurements (Fig.~\ref{fig:fig01}) during those months. One interval was chosen arbitrarily for each year between 2003 and 2008, in which the measurements had no data gap (See Table~\ref{tab:intervalls}). The durations of the intervals are between 50\,min and 28\,hours. We also used Cluster Ion Spectrometer Hot Ion Analyzer (CIS/HIA) spin averaged resolution data \cite{reme01:_first_clust_cis} on Cluster SC1 to determine the solar wind speed. We used this parameter to transform the data into the solar wind reference frame, which was the first step of the correlation calculation.

\section{Multi-spacecraft methods}
\label{sec:tech}

\begin{figure}
\centerline{\includegraphics[width=0.5\textwidth,clip=]{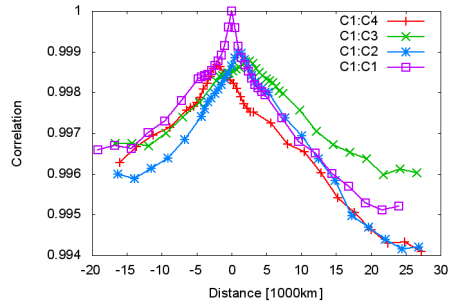}
\caption{An example of the measured sections of the correlation function. $Ci$ is the $i^{th}$ Cluster spacecraft. The red line (C1:C4) for example shows the correlation of the magnetic data measured on Cluster1 and Cluster4.}\label{fig:fig02}}
\end{figure}

The purpose of this study is to determine full 3D correlation functions. Using the 1s averaged FGM measurement we calculated one dimensional time-lagged auto- and cross-correlation functions of the magnetic fluctuations (Cluster C1-C1, C1-C2, C1-C3, C1-C4, See Fig.~\ref{fig:fig02}) in every ten minutes of the measurement intervals of Table~\ref{tab:intervalls}. These are one dimensional sections of the 3D correlation function. After the calculation of the sections we fitted a single 3D test function to all of the points of all four sections. 

The correlation lengths are quite long in the solar wind but the separation of the Cluster fleet was enough to measure significant variations, therefore to calculate the correlations. Every section was calculated in a 120s long interval, i.e. the time lags was less then 60 seconds. Adaptive correlation calculation was used to decrease the calculation time:
\begin{enumerate}
\item First a draft of the correlation function was calculated by 5\,s long leapfrog steps. 
\item The maximum of the correlation functions was determinated.
\item Finally the correlation function was calculated 10 s before and after the maximum using 1s resolution.
\end{enumerate}
This method spared computational resources and focused on the most important (most rapidly changing) part of the correlation function. 

\begin{figure}
\centerline{\includegraphics[width=0.5\textwidth,clip=]{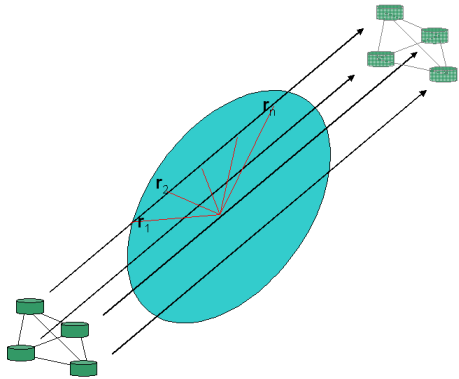}
\caption{Illustration of our method: The four Cluster spacecraft sampling a volume of the solar wind. We compute the correlations along $\mathbf{r}_1$-$\mathbf{r}_n$, and fit a 3D correlation function to the resulting data.}\label{fig:fig03}}
\end{figure}

Using the Taylor's or ``frozen in field'' hypothesis \cite{taylor38:_spect_of_turbul}, these one dimensional, time dependent measurements can be translated to samples of the 3D spatial correlation function along a multitude of $\mathbf{r}_n$ space vectors (Fig.~\ref{fig:fig03}) in solar wind frame.

The Taylor's hypothesis states that on a time scale short compared to the characteristic timescale on which the fluctuations vary in the solar wind frame, the variations measured by the spacecraft are due to the convection, i.e. approximately static fluctuations are convected past the spacecraft. This is true only if the convection speed is much greater than any other speed influencing the measurements. Since both the velocity of the spacecraft and the wave propagation speed (roughly the Alfv\'en speed $V_A$) are quite low compared to the solar wind speed ($V_{sw}$) at 1\,AU, the time lag ($t$) recorded by a spacecraft can be converted to a spatial position as $\mathbf{x}=\mathbf{V}_{sw}t$ \cite{taylor38:_spect_of_turbul}. 

The Taylor's hypothesis however has a modified form for multi-spacecraft case. In this case the distance of two measurement positions in the solar wind frame can be quite low (even zero), thus the finite wave propagation speed becomes significant \cite{horbury00:_clust_ii_analy_of_turbul}. In this case the condition for the Taylor's hypothesis can be written as 
$ \left| \mathbf{r}_{1,2} - \mathbf{V}_{sw}t \right| \gg \left| {\mathbf{V}_A t} \right| $ where $\mathbf{r}_{1,2}$ is the distance of the two spacecraft. The condition is fulfilled for the vast majority of our measurement points. For the few cases, when it is not, we have only a single corrupted point out of the 160 we use to fit a correlation function. ($ \left| \mathbf{r}_{1,i} - \mathbf{V}_{sw}t \right| $ could only be small for one point for fixed $\mathbf{r}_{1,i}$ vectors.) This single point has no significant influence on the fit.

\begin{figure}
\centerline{\includegraphics[width=0.5\textwidth,clip=]{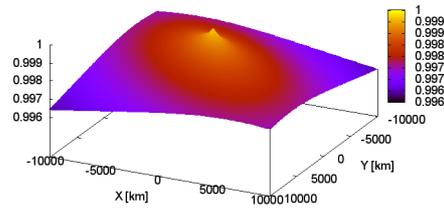}
\caption{A two dimensional section of the 3D correlation function that we used to fit the data. The correlation values are given on the vertical axis and also by a color code with respect to the spacecraft separation in the X and Y directions}\label{fig:fig04}}
\end{figure}

We can now fit a 3D test function (an Ansatz) C(x,y,z) to these samples (Fig.~\ref{fig:fig04}). The quality of the fit is measured by the sum of the squared deviations of our function and the measured values. We found that the function 
\begin{eqnarray*}
\label{eq:eq1}
C\left(x,y,z\right)=\exp\left(-\left[\frac{{x'}^2}{a^2}+\frac{{y'}^2}{b^2}+\frac{{z'}^2}{c^2}\right]^{\gamma}\right)  
\end{eqnarray*}
fits very well in most of the time intervals we have studied. (Here a, b and c are the semi-principal axes of the correlation ellipsoids; x', y' and z' are transformed into the principal coordinate system.) 

Random processes can usually be described by an exponential correlation function $ C(r)\propto e^{-\frac{\left|r\right|}{\lambda}}$, where $\lambda$ is the correlation length. Actually the correlation length is defined by this equation. (When the correlation function does not show exponential decay, the correlation length is infinite; these are the so-called strongly correlated systems, occurring usually only in very special situations such as phase transitions, critical points and alike.)

When we set the $\gamma$ parameter to be $\gamma=1/2$ in our correlation function $C\left(x,y,z\right)$, then that is only a 3D version of the usual equation, with three independent correlation lengths. Unfortunately if we fix $\gamma$ to a certain value, the quality of the fit is significantly worsened, so we were forced to use $\gamma$ as an independent parameter.

When $\gamma \ne 1/2$ the correlation function is a stretched-exponential. (This kind of correlation is often found in systems with broad distribution of correlation scales.) In this case the semi principal axes of the correlation ellipsoids are not exactly the classical correlation lengths, but their ratios serve as a very good measure of the anisotropy.

We have seven independent parameters in a transcendent function (a, b, c, $\gamma$, and the three independent parameters of the rotation matrix)  so we encountered serious problems during our analysis:  
\begin{itemize}
\item The Ansatz is a transcendent function so linear regression cannot be used. 
\item The exponent is unknown so it cannot be solved for the logarithm. 
\item The rotation matrix has 9 parameters but only 3 of them are independent. 
\end{itemize}
So our task was to fit a transcendent function with 9+3+1 parameters and 6 constraints.

We solved this task by using the least squares method in a special way. We re-defined the Ansatz by using a special parameter set taking into account the constraints. The new parameter set eliminates the usage of the transformation matrices in the parameterization of the correlation ellipsoids:
\begin{eqnarray*}
\label{eq:eq2}
C=\exp\left(-A\cdot 10^{-5}\left[\left(x+By\right)^2+\left(Cx+Dz\right)^2+\left(Ey+Fz\right)^2\right]^{\gamma}\right) 
\end{eqnarray*}
Then we let the system move in the 7 dimensional parameter space on a surface defined by the least squares toward the minimum of the surface. Some random step was forced to avoid getting the local minima instead of the real minimum. Adaptive step size was used to spare resources and focus on the important region. Applying this method with these parameters the code finds stable minimum most of the time. The minimum does not depend on the choice of initial parameters. The final value of the error sum is 3-5 orders of magnitude smaller than its initial value. This value shows how well $C(x,y,z)$ fits to the 160-160 measurement points of each evaluations. The sum is usually in the $10^{-6} - 10^{-5}$ range, which means that the error for a single measurement point is approximately $1 - 3 \times 10^{-4}$. Since it is really difficult here to estimate the error from basic principles (among other things it depends on the relative positions of the spacecraft and the physical properties of the plasma) we used this error sum of the least squares fit to estimate the overall error of our method.

Though the new parameters are extremely useful in solving the fitting problem, it is not easy to find the physical meaning of them. But after the fit we can reconstruct the physical parameters for further analysis. From our special parameters we compute the magnitudes and directions of the semi principal axes of the correlation ellipsoids for $C\left(x,y,z\right)=0.99$.

\section{Discussion  and conclusion}
\label{sec:disc}

\begin{figure}
\centerline{\includegraphics[width=0.8\textwidth,clip=]{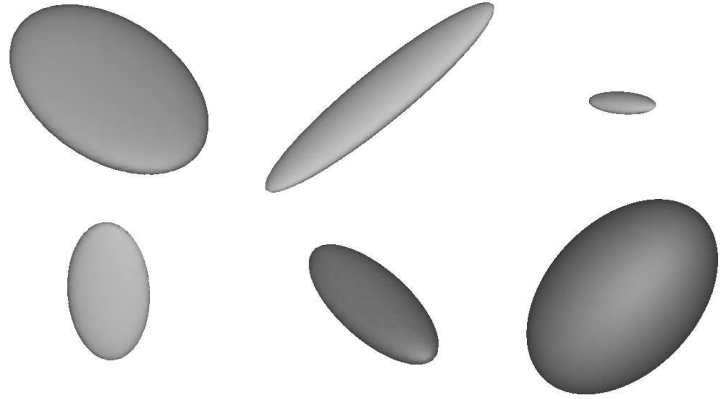}
\caption{Examples of the correlation ellipsoids described in Sec.~\ref{sec:tech}. The ellipsoids are the ''equipotential'' surfaces for a fixed ($C\left(x,y,z\right)=0.99$) value. The units are arbitrary but the size of the object is proportional. The correlation function is clearly essentially anisotropic.}\label{fig:fig05}}
\end{figure}

A lot of ellipsoids were visualized given by the algorithm described in Sec.~\ref{sec:tech}. These ellipsoids show significant variations in size as well as shape. Most of them show strong three axes anisotropy. This means that the correlation functions and so the fluctuations of the solar wind plasma are basically anisotropic (Fig.~\ref{fig:fig05}) they have neither spherical nor axial symmetry. 

\begin{figure}
\centerline{\includegraphics[width=0.5\textwidth,clip=]{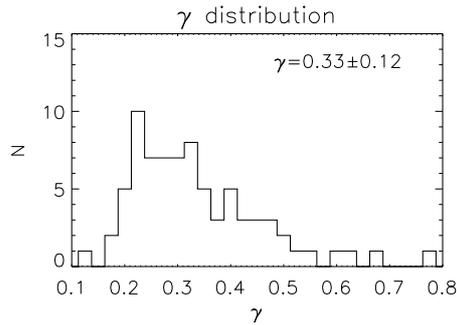}
\caption{The distribution of the $\gamma$ exponent of the stretched-exponential correlation function (see Sec.~\ref{sec:tech}), with its mean value and standard deviation. N is the number of evaluations with $\gamma$ values falling in the corresponding 0.025 wide bin.} \label{fig:fig06}}
\end{figure}

The $\gamma$ exponent of the stretched-exponential correlation function shows a broad distribution. Its mean value and standard deviation are 0.33 and 0.12, respectively (Fig.~\ref{fig:fig06}).

\begin{figure}
\centerline{\includegraphics[width=0.3\textwidth,clip=]{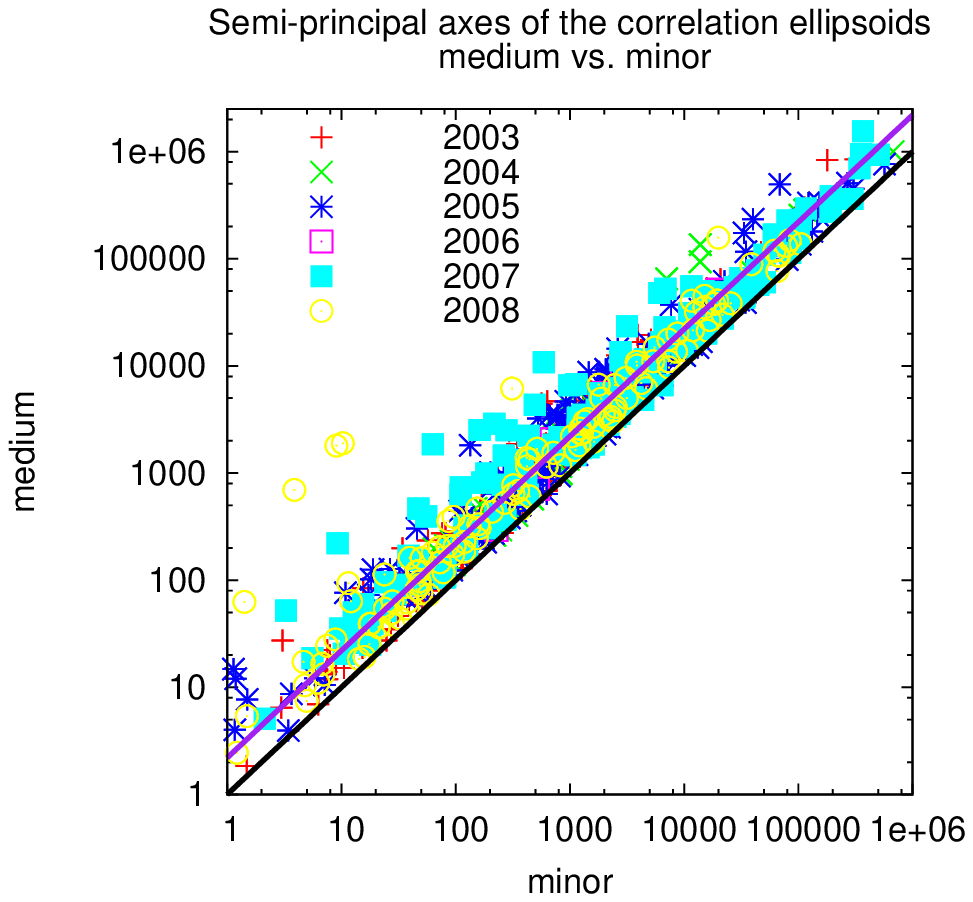}
\includegraphics[width=0.3\textwidth,clip=]{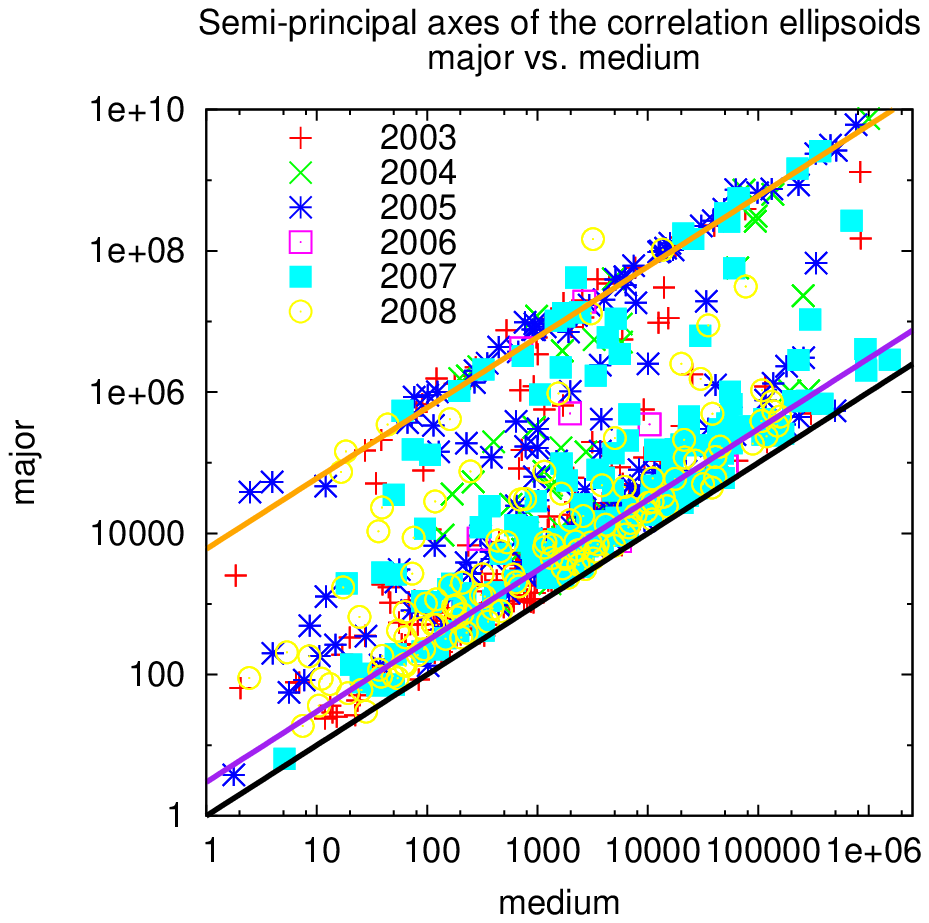}
\includegraphics[width=0.3\textwidth,clip=]{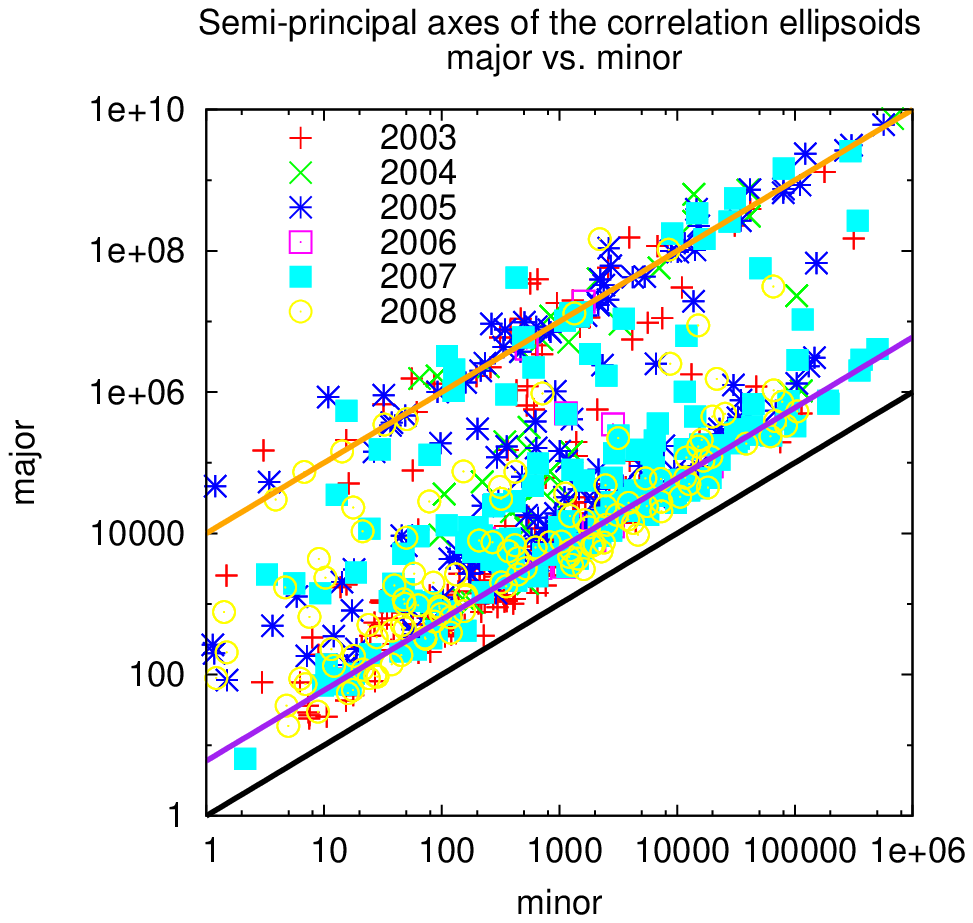}
\caption{Ratios (c:b, c:a, b:a) of the semi-principal axes of the correlation ellipsoids (i.e. ratios of the anisotropic correlation lengths). Samples from year 2003 to 2008. Black line: ratio=1. Purple line: fit for the first population. Orange line: fit for the second population.}\label{fig:fig07}}
\end{figure}

\begin{figure}[t]
\centerline{\includegraphics[width=0.3\textwidth,clip=]{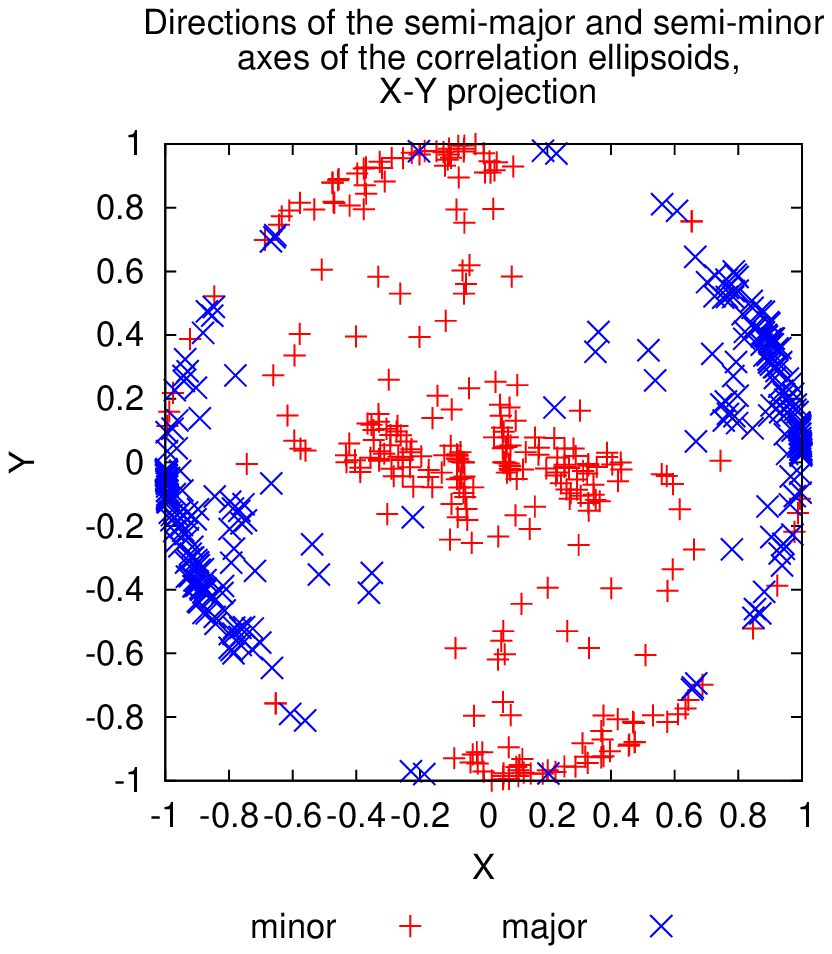}
\includegraphics[width=0.3\textwidth,clip=]{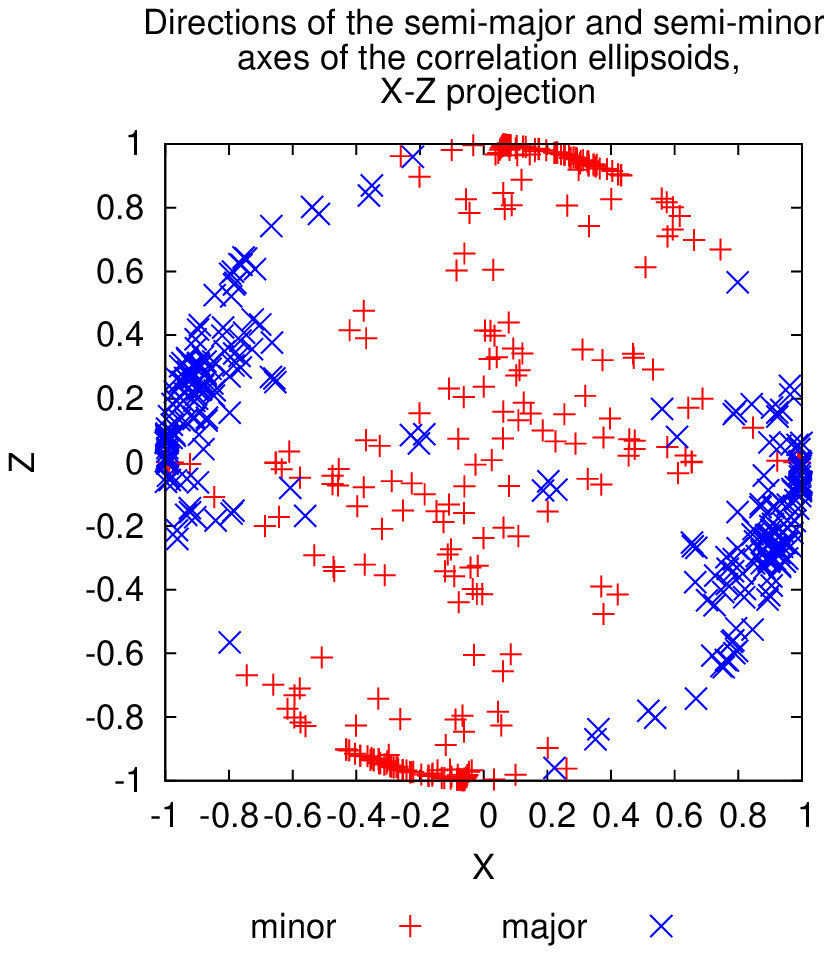}
\includegraphics[width=0.3\textwidth,clip=]{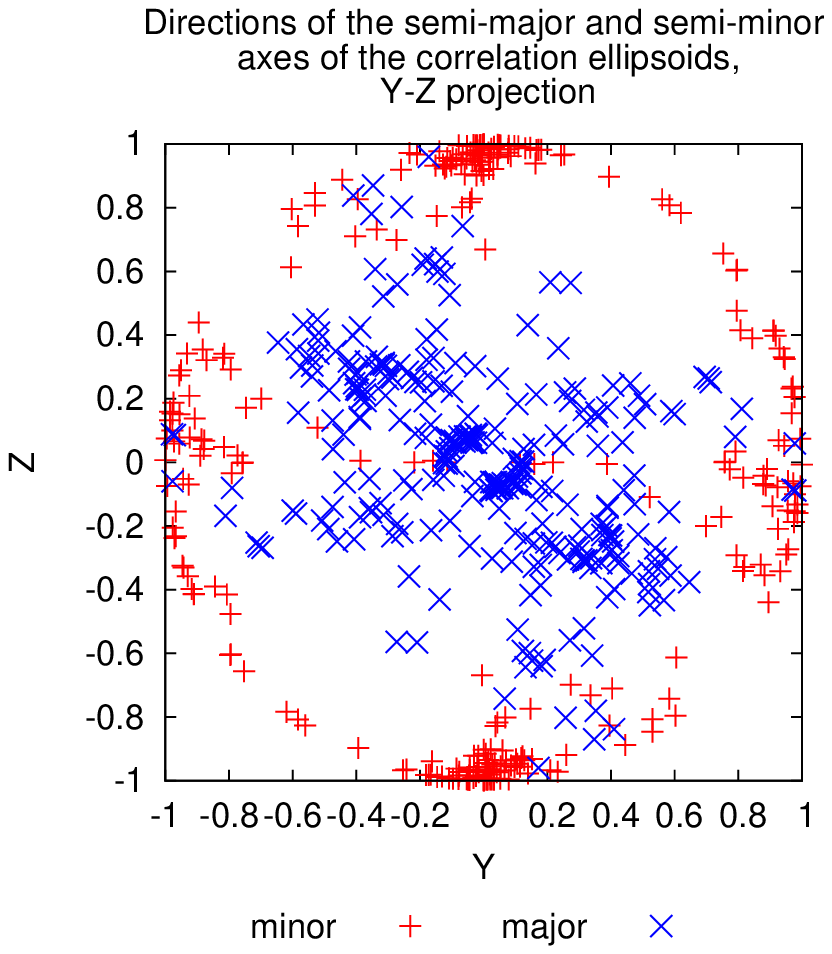}
\caption{The directions of the three principal axes of the correlation ellipsoids. GSE coordinates, (a) X-Y, (b) X-Y and (c) Y-Z projections. Computed from data measured on 04/03/2005. Red: minor axes, blue: major axes.}\label{fig:fig08}}
\end{figure}

We found that small scale IMF fluctuations are highly inhomogeneous: the correlation length varies over almost six orders of magnitude. The IMF turbulence shows significant anisotropy with two distinct populations. Most of the time the ratio of the 3 principal axes of the correlation ellipsoid, i.e. the ratio of the correlation lengths, is roughly  1:2.2:6, in the remaining time the major axes can be much larger (Fig.~\ref{fig:fig07}). In this second population the ratio of axes shows high variations but the ratio 1:2.2:60000 has slightly higher probability. This means extremely strong correlation in a given direction. We found favoured directions in the orientation of the correlation ellipse (Fig.~\ref{fig:fig08}). The major axes favor the Sun-Earth direction. The minor axes favour the direction perpendicular to the ecliptic plane. In most of the time, the correlation lengths computed using $B_z$ data (GSE coordinates) are significantly smaller than those computed using $B_x$ or $B_y$. The cause of this could be the relative lack of slab turbulence. 

\begin{ack}
The work of G. Facsk\'o was supported by the OTKA grant K75640 of the Hungarian Scientific Research Fund. The authors thank Iannis Dandouras (CIS PI) and the Cluster Active Archive for providing CIS HIA solar wind measurements and Anna-M\'aria V\'\i gh for improving the English of the paper. 
\end{ack}

% References ----------

\bibliographystyle{spr-mp-sola}
\bibliography{sola-s-10-00092-4-tx}

\end{article}

\end{document}